\documentclass[manuscript]{emulateapj}
\usepackage{amsmath}

\shorttitle{FRB every second}
\shortauthors{Fialkov \& Loeb}
\usepackage{color}

\begin{document}

\title{A Fast Radio Burst Occurs Every Second throughout the Observable Universe}

\author{Anastasia Fialkov \& Abraham Loeb}
\affil{Harvard-Smithsonian Center for Astrophysics, Institute for Theory and Computation, 60 Garden Street, Cambridge, MA 02138, USA}
\email{anastasia.fialkov@cfa.harvard.edu \\ aloeb@cfa.harvard.edu}

\begin{abstract}
  Recent multi-telescope observations of the repeating Fast Radio Burst FRB 121102 reveal a Gaussian-like spectral profile and associate the event with a dwarf metal-poor galaxy at a cosmological redshift of 0.19. Assuming that this event represents the entire FRB population, we make predictions for the expected number counts of FRBs observable by future radio telescopes between  50 MHz and 3.5 GHz. We vary our model assumptions to bracket the expected rate of FRBs, and find that it exceeds one FRB per second per sky when accounting for faint sources.  We show that future low-frequency radio telescopes, such as the Square Kilometer Array, could detect more than one FRB per minute over the entire sky originating from the epoch of reionization.  
\end{abstract}

\keywords{cosmology: theory}

\section{Introduction}

Fast radio bursts (FRBs) are rapid (millisecond duration) and bright ($\sim $Jy) radio transients with excess dispersion measure (DM) beyond the maximum column of Galactic electrons along the line of sight. Only 23 FRBs has been observed so far out of which 17 were detected with the Parkes Radio Telescope at 1.4 GHz frequency \citep{Lorimer:2007, Keane:2011, Thornton:2013, Burke-Spolaor:2014, Petroff:2015, Ravi:2015, Champion:2016, Keane:2016,  Ravi:2016,  Petroff:2017}, one  with the  Green Bank Telescope (GBT) in the 700-900 MHz frequency band  \citep{Masui:2015}, one in the 1.4-GHz Pulsar ALFA survey with the Arecibo Observatory \citep{Spitler:2014}, three at 843 MHz with UTMOST \citep{Caleb:2017}, and one with the Australian Square Kilometre Array Pathfinder   \citep[ASKAP,][]{Bannister:2017} in the range between 0.7 and 1.8 GHz (see the online FRB catalog\footnote{http://www.astronomy.swin.edu.au/pulsar/frbcat/} for more details on the detected events).

Out of the observed FRBs, the one detected by Arecibo, tagged FRB 121102,  is the only event found to repeat \citep{Spitler:2016}, although see \citet{Piro:2017}. The repeating signal made it  possible to pinpoint the location of the source on the sky to within 0.1 arcsec \citep{Chatterjee:2017} and associate it with a dwarf galaxy  at $z=0.19$ \citep{Tendulkar:2017, Bassa:2017}. This groundbreaking discovery showed that  FRBs  (or at least FRB 121102) are extra-galactic transients  that could be used for cosmology.  The host galaxy was  identified  to be metal poor (with the observed metallicity of $12+\log(O/H) \lesssim 8.4$), have stellar mass of $M_* = \left(4-7\right)\times 10^7$ M$_\odot$, a star formation rate of 0.4 M$_\odot$ yr$^{-−1}$ and a substantial host dispersion measure of $\lesssim 324$ pc cm$^{-−3}$. 

Recent multi-frequency observations of FRB 121102 with multiple  radio telescopes  including the Karl G. Jansky Very Large Array (VLA, $2.49-3.51$ GHz), Arecibo (1.4 GHz), Effelsberg (4.85 GHz), the  Long Wavelength Array (LWA1, $52.2-–71.8$ MHz and $68.2–-87.8$ MHz)  and the Arcminute Microkelvin Imager Large Array (AMILA, $13-18$ GHz) yielded simultaneous detection by Arecibo and the VLA bands; while LWA,  Effelsberg and AMILA did not measure any signal \citep{Law:2017}. The results show that the bursts with simultaneous observing coverage are not well described by a power-law flux density model.  The VLA observations indicate that the spectrum of each repeating burst has a bell-like  profile and can be fitted with a Gaussian profile, with best fit parameters varying from one repetition to another. The central frequency of the profile varies between $\nu_c = 2.8$ and 3.2 GHz, the peak fluxes of the bursts fall in the range of $130-3340$ mJy, and the FWHM  is in the range of $290-690$ MHz. 

The origin of FRBs is still a mystery with possible explanations including young, rapidly spinning magnetars   \citep[e.g.,][]{Metzger:2017, Beloborodov:2017, Cordes:2016}. The nature of FRBs should affect their event rates,  luminosity and host galaxy demographics \citep[e.g.,][]{Nicholl:2017}. However, due to the scarcity of observations, these properties of the population are poorly constrained  \citep{Vedantham:2016, Li:2016, Lawrence:2016}. For instance, the fluence-complete rate above a fluence of $\sim 2$ Jy ms was shown to be $2.1^{+3.2}_{−1.5} \times 10^3$ FRBs sky$^{-−1}$ day$^{-−1}$ \citep{Keane:2015}. \citet{Nicholl:2017} compared the volumetric rates and host galaxy demographics of  millisecond magnetar remnants of superluminous supernovae (SLSNs) and gamma ray bursts (GRBs) to the observed FRBs, showing that, if FRB 121102 is typical, properties of the observed ensemble  are consistent with expectations for millisecond magnetars. If this applies to the entire FRB population, most FRBs are probably repeaters  because otherwise their birth rate would be too high compared to the rate of SLSNs.

Given the scarcity of observations, it is prudent to adapt  a phenomenological approach when modeling a cosmological population of FRBs. For instance, \citet{Caleb:2016} assume flat spectral index of FRBs and a log-normal luminosity function (LF), and examine whether the properties of observed events are consistent with such a cosmological population. However, the Gaussian-like spectral profile of FRB 121102 in the observed band indicates that the spectrum for at least this event is not well modeled as a power law. If all FRBs exhibit emission in a limited bandwidth,  distant sources would redshift out of the observed band, affecting predictions for a cosmological population, and thus lower frequency surveys would be necessary to detect high-redshift FRBs. Currently we do not know if there is any signal outside of the observed frequency range, and if FRB 121102 is representative.

In this Letter we predict expected numbers  of FRBs as a function of telescope sensitivity and frequency band  comparing models with a flat spectrum and a Gaussian-like spectrum.  We  vary our prescription for the LF and the typical mass of host galaxies to provide upper and lower limits on the expected rate of FRBs.   The paper is organized as follows: in Section \ref{Sc:Model} we outline our assumptions regarding the population of host galaxies, FRB luminosity and intrinsic rates. In  Section \ref{Sc:Res} we compare our model with the existing data and make predictions for future more sensitive surveys. We summarize our conclusions in Section \ref{Sc:Conc}.

\section{Modeling the population}
\label{Sc:Model}


\subsection{Host Galaxies}

The fact that FRB 121102  resides in a metal-poor dwarf galaxy indicates that transients of its type do not trace star formation at all metallicities (in which case a typical FRBs would occur in more massive galaxies).  It might also be an indication that FRBs are more common in the  high redshift Universe which is filled with metal-poor low-mass galaxies\footnote{If FRBs indeed occur mostly in  small  metal-poor galaxies, they  resemble  long GRBs and hydrogen-poor SLSNs \citep{Modjaz:2008, Lunnan:2014, Nicholl:2015}.}. 

 We consider two prescriptions for the hosts. First, we assume that all host galaxies have similar properties to the one in which FRB 121102 is located. In particular, we require galaxies to have  stellar mass in the range between  $M_{*,1} = \overline{M}_*/\sqrt{10}$ and $M_{*,2}=  \overline{M}_* \sqrt{10}$, where $\overline{M}_* = 5.3\times 10^7$ M$_\odot$  is the stellar mass measured for FRB 121102. The observed metallicity of the  host galaxy of FRB 121102 agrees well with theoretical predictions for galaxies of its mass and does not add any constraints \citep{Ma:2016, Hunt:2016}. Second, we assume high-mass hosts with $\overline{M}_* = 5\times 10^9$ M$_\odot$ for comparison. Such galaxies are much more common in the low-redshift Universe than the metal poor dwarfs.  

The number of host galaxies per comoving volume with stellar masses spanning the range between  $M_{*,1}$ and $M_{*,2}$ is set by the number of dark matter halos that contain such galaxies and can be computed using the excursion set formalism. To infer this number we: (i) use  Sheth-Tormen prescription  \citep{Sheth:1999} for  the number density of dark matter halos versus halo mass, $dn/dM_h$, and (ii) populate halos by stars using abundance matching and a star formation efficiency which varies with halo mass and redshift \citep{Behroozi:2013}. We select halos hosting galaxies with stellar mass in the range between  $M_{*,1}$ and $M_{*,2}$ as FRB hosts, yielding the  observed number of FRBs,  per unit time
\begin{equation}
N_{FRB} =  \int_V \int_{M_h = M_h(M_{*,1})}^{M_h = M_h(M_{*,2})}\frac{dn}{dM_h}\frac{R_{int}}{(1+z)} dM_hdV, 
\label{Eq:NFRB}
\end{equation}
where $dV$ is the comoving volume element, the factor $\left(1+z\right)$ accounts for the  time dilation effect, and $R_{int}$ is the intrinsic rate of FRBs produced  by each galaxy  in the direction of observer, into which we absorb  the  birth rate of each emitter, the lifetime of an emitter, the rate at which each source emits repeated FRB signals, and beaming factor (see Section \ref{Sec:rint} for details). In its current form, Eq. (\ref{Eq:NFRB}) assumes that all produced FRBs are observable. When calculating the number of observed FRBs with peak fluxes, $S_{peak}$, above a given flux limit, $S_{lim}$, we  modify the equation to include  a  selection rule $S_{peak}>S_{lim}$ with $S_{peak}$ dependent on the redshift and  the intrinsic FRB luminosity, as discussed in the next subsection.

\subsection{FRB Spectra}
\label{Sec:LF}
While the bright end of the FRB population is somewhat constrained by existing observations, the faint end is highly unconstrained and will be probed by future sensitive telescopes. We consider several options for assigning a luminosity to each event, thus bracketing a wide range of possible LFs for the entire FRB population:\\
{\bf Model A (the lower limit on faint FRBs):} FRBs are standard candles with all events having the same intrinsic luminosity, $L_{int}$, and a Gaussian-like spectral profile.  Using FRB 121102 as a prototype, we construct an intrinsic spectral profile with the observed flux fitted by
\begin{equation}
S_{obs}(\nu_{obs}) = S_{peak}\exp\left[-\frac{(\nu_{obs}-a)^2}{2c^2}\right].
\label{Eq:SED}
\end{equation}
We normalize $L_{int}$ to yield $S_{peak} = 0.9017$ Jy for a source at $z = 0.19$ adopting $a = 2.986$ GHz, and $c = 0.1984$ GHz \citep[mean values for the repeater, see Table 2 of][]{Law:2017}. This model gives a delta LF and provides a lower limit on the number of faint FRBs which are those located at large cosmological distances.  \\
{\bf  Model B:} to investigate the effect of the spectral  shape we compare the results of Model A to a case with a flat spectrum normalized to the same value of $S_{peak}$. \\
{\bf Model C (the upper limit on the faint FRBs):} each event has an average Gaussian-like profile as in  Model A, but the LF has the Schechter form
\begin{displaymath}
\frac{dN}{dL} \propto \left(\frac{L}{L_\star}\right)^{\alpha}e^{-L/L_\star},
\end{displaymath}
where $L_\star$ is set so that FRB 121102 will match the average luminosity at $z=0.19$ (with no dependence of $L_\star$ on redshift). We integrate over the LF when calculating the number of observed FRBs from Eq. (\ref{Eq:NFRB}). This model gives the maximal possible number of intrinsically faint FRBs for  $\alpha = -2$, the steepest slope for which the luminosity density does not diverge. The observed faint population includes both the intrinsically faint sources and those at large cosmological distances.\\
{\bf Model D:} FRBs have a flat spectrum as in Model B, but their LF has the Schechter form.

\subsection{Intrinsic FRB Rate}
\label{Sec:rint}

Assuming a constant intrinsic FRB rate per galaxy, we can find $R_{int}$ in Eq. (\ref{Eq:NFRB}) by matching the predicted $N_{FRB}$ to available observations. The projected all sky FRB rate $R_p = 2\times 10^3$ sky$^{-1}$ day$^{-1}$ measures the total number of events observed from  $z \lesssim 1$   with a fluence sensitivity of 1 Jy ms \citep{Law:2017, Nicholl:2017} and yields a different value of $R_{int}$ for each one of the considered models. 
To extract $R_{int}$ from Eq. (\ref{Eq:NFRB}) we assume a  survey with a flux limit of $S_{lim} =1$ Jy observing all the events out to $z = 1$ (assuming a 1 ms duration).   Naturally, the inferred rate depends on the properties of the host galaxies. In the case of low-mass galactic hosts, the intrinsic rate  is $R_{int} \approx 2\times 10^{-5}$ per galaxy per day for Models A and B  and $R_{int} \approx 3\times 10^{-3}$ for  Models C and D. The normalization in the case of high-mass hosts is $R_{int} \approx 3\times 10^{-4}$  for Models A and B and $R_{int} \approx 4\times 10^{-2}$    for Models C and D. 

If more details on the nature of FRBs were known, their intrinsic rate, could be inferred from the relation: $R_{int} =  R_{b} \tau R_{s} \Omega_b $, where $R_{b}$ is the birth rate of FRB emitters, $\tau$ is their lifetime,  $R_{s}$ is the rate at which each source emits repeated FRB signals, and $\Omega_b$ is the beaming factor.

\section{Results}
\label{Sc:Res}

\subsection{Matching Existing Observations}

\begin{figure}
\begin{center}
\includegraphics[width=2.4in]{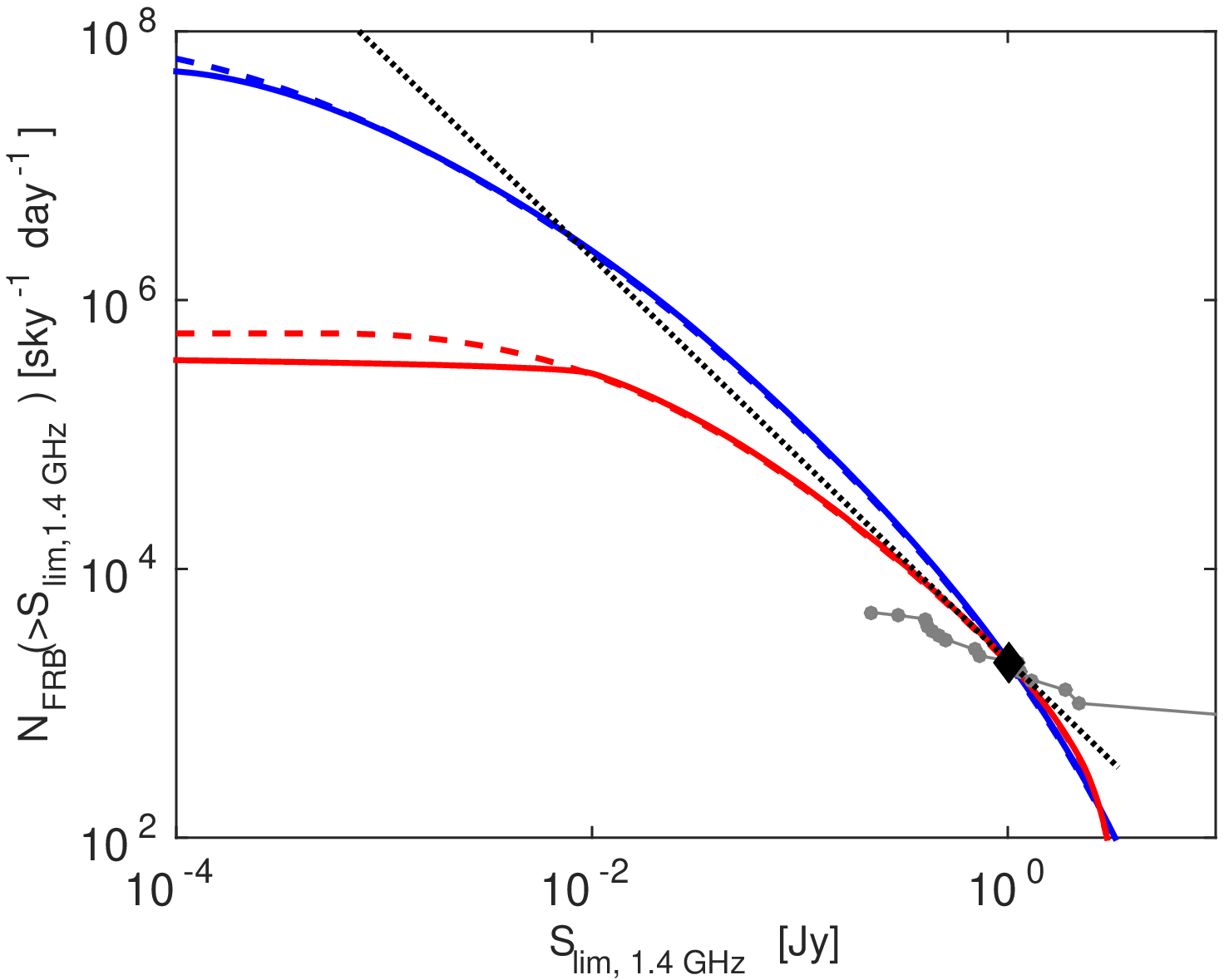}
\includegraphics[width=2.4in]{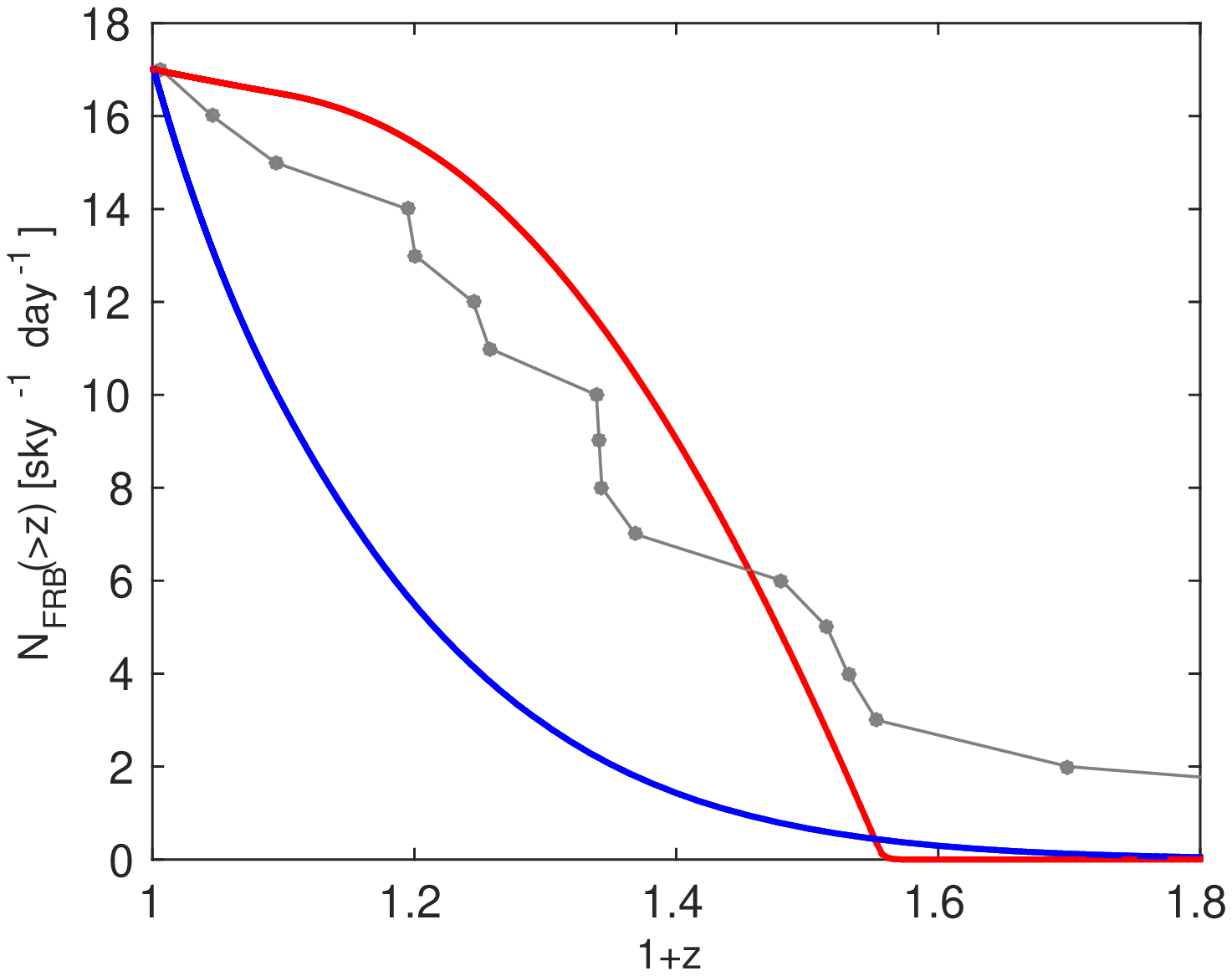}
\end{center}
\caption{{\bf Top:} FRB detection rates  at 1.4 GHz ($1.25-3.5$ GHz band). We show the results for Models A (red solid), B (red dashed), C (blue solid), and D (blue dashed), assuming low-mass host galaxies. The case with high-mass hosts looks identical at 1.4 GHz since  $R_{int}$ varies, as indicated in the text. We also show the case of  a uniform distribution of sources  with $N_{FRB} \propto S^{-3/2}$ (dotted). The black diamond shows the normalization of $2\times 10^3$ sky$^{-1}$ day$^{-1}$ \citep{Law:2017}, the grey circles show the cumulative number counts of the observed events deduced from measurements with Parkes, ASKAP and Arecibo. {\bf Bottom:} Expected number of observed FRBs as a function of redshift for $S_{lim} = 0.1$ Jy with $DM_{host}=324$ pc cm$^{-3}$, shown  for Models A (red solid) and  C (blue solid), assuming low-mass hosts. The grey circles mark redshifts of the FRBs observed by Parkes, ASKAP and Arecibo and are inferred  from  the data  assuming fully ionized IGM. }   
\label{Fig:LF1}
\end{figure}

We compare model predictions to the surveys conducted by Parkes,  Arecibo and ASKAP (which found  19 different FRBs in total) based on the online FRB catalog. We ignore  FRBs observed by GBT  and UTMOST because they probe a different frequency band (0.8 GHz compared to 1.4 GHz). The top panel of Figure  \ref{Fig:LF1}  shows the number counts of FRBs observed at 1.4 GHz as a function of flux limit in the band. We rescale the observed numbers to match the total rate of  $2\times 10^3$ sky$^{-1}$ day$^{-1}$  at $S_{lim}=1$ Jy. As shown in  Figure  \ref{Fig:LF1}, the predicted FRB number counts in all models, as well as in the case where the sources are  uniformly distributed in comoving volume (dotted line) overpredict the observed numbers at the faint end and underpredict them  at the bright end. For all models, including Model A which is designed to yield the lower limit on faint FRB numbers,  our curves are steeper than observed.

The discrepancy between our models and observations might indicate that above the flux limit the surveys  of faint FRBs are not complete, and miss many events;  while  the bright end is most likely highly biased. An additional possibility is that low-luminosity FRBs do not exist or that the  LF has a positive spectral index (which, however, is not typical). The deficit of FRBs at low frequencies might also be caused by self absorption around the source if the line-of-sight goes through the emitting region of the quiescent radio source \citep{Metzger:2017}. \citet{Cordes:2017} argued that if a burst passes through a galactic disk where the gas density is large enough, it could be scattered into non-detectability.  However, the probability of having such a small impact parameter from an unrelated galactic center along a random line of sight through the Universe is very small even for sources at $z\sim 10$. Such events would also be strongly lensed, and \citet{Barkana:2000} have shown that the lensing optical depth is still much smaller than unity even for high redshift sources.

The expected numbers  rise at low fluxes until saturate. Our models predict that including all  faint events (high-redshift events in Models A and B and high-z plus low-luminosity events in Models C and D), we expect more than $3.5\times 10^5$ events per day from the entire Universe (i.e., more than one burst per second). 

The Gaussian spectral profile  leads to  a factor of $20-40\%$ deficit in the expected number of faint events at 1.4 GHz and $S_{lim} = 0.1$ mJy, compared to a flat spectrum (Model A vs. B and Model C vs. D). This is because if the spectrum is line-like, emission from distant sources  redshifts out of the band and is no longer observable. This effect is much stronger at lower frequencies. 

Comparing Model A and C we find that the Schechter LF yields many more faint sources. Note, however, that one needs a much higher $R_{int}$ in this case to match the low-redshift observations because a smaller fraction of the events have high observed flux.

The cumulative number of observed FRBs with $S_{lim}=0.1$ mJy and redshifts above $z$ is shown on the bottom panel of the Figure \ref{Fig:LF1}. For each observed FRB  we compute the redshift of the host galaxy by subtracting from the measured DM the contribution of the Milky Way \citep[NE2001 model of][]{Cordes:2002} and assuming fully ionized IGM. Based on the properties of FRB 121102 we  assume that the contribution of each host to DM is  $ 324$ pc cm$^{-3}$ \citep{Tendulkar:2017}.  Because of the relatively high Galactic contribution, the cosmological DM for 2 out of 19  FRBs (010621 and  150807) results in negative DM values. We exclude these 2 events, which leaves 17 FRBs in total. We re-normalize our models to yield this number of events. 

Figure \ref{Fig:LF1} shows that once FRBs will be observed at greater numbers, the redshift evolution of their number counts can differentiate between the Schechter LF and the standard candle scenarios. The drop with redshift is more significant in the case of the Schechter LF,  because a large fraction of the sources are intrinsically faint and quickly redshift below the flux detection limit even if their central frequency is still within the telescope band. 

\subsection{Low-Frequency Observatories}

\begin{figure*}
\begin{center}
\includegraphics[width=2.2in]{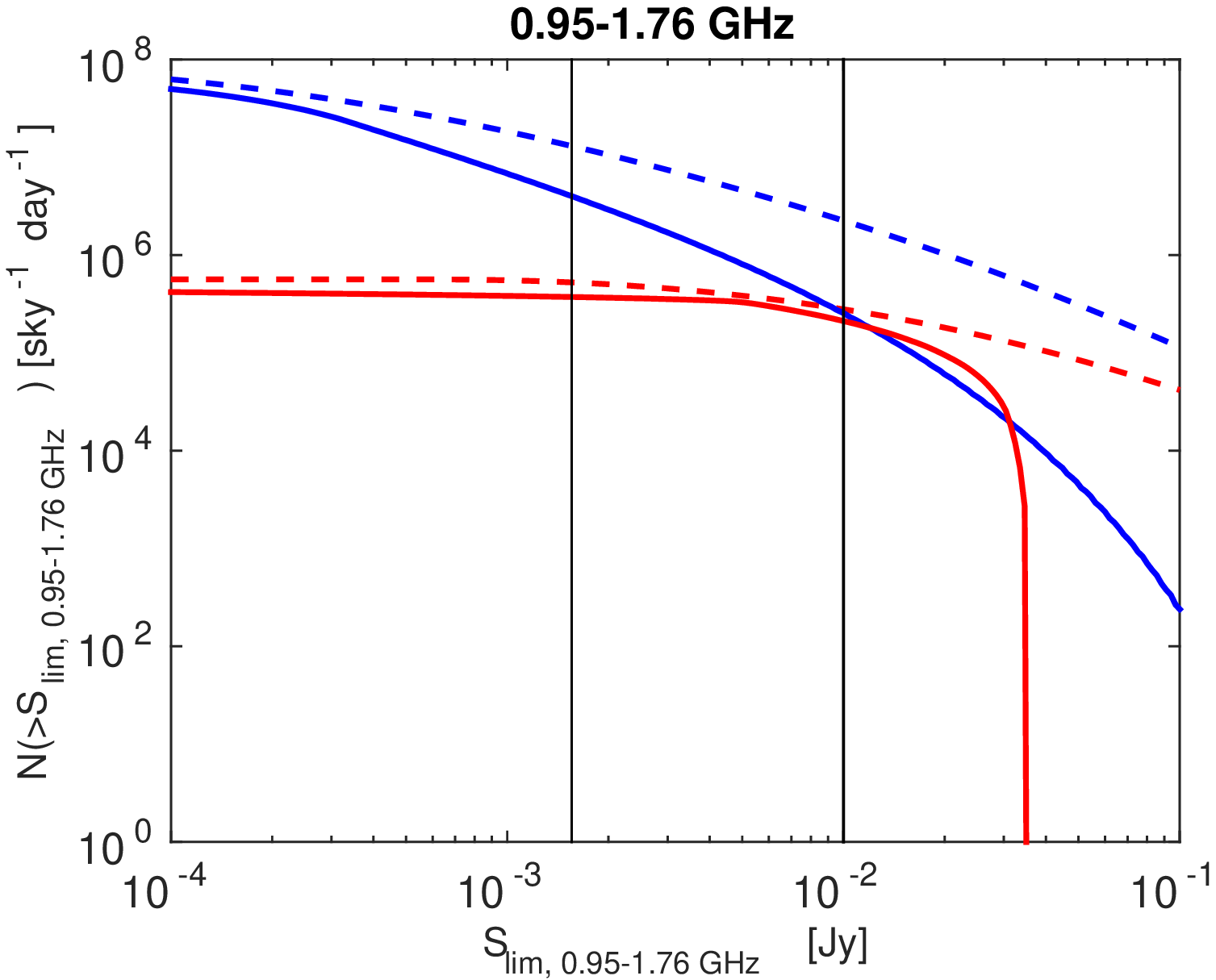}\includegraphics[width=2.2in]{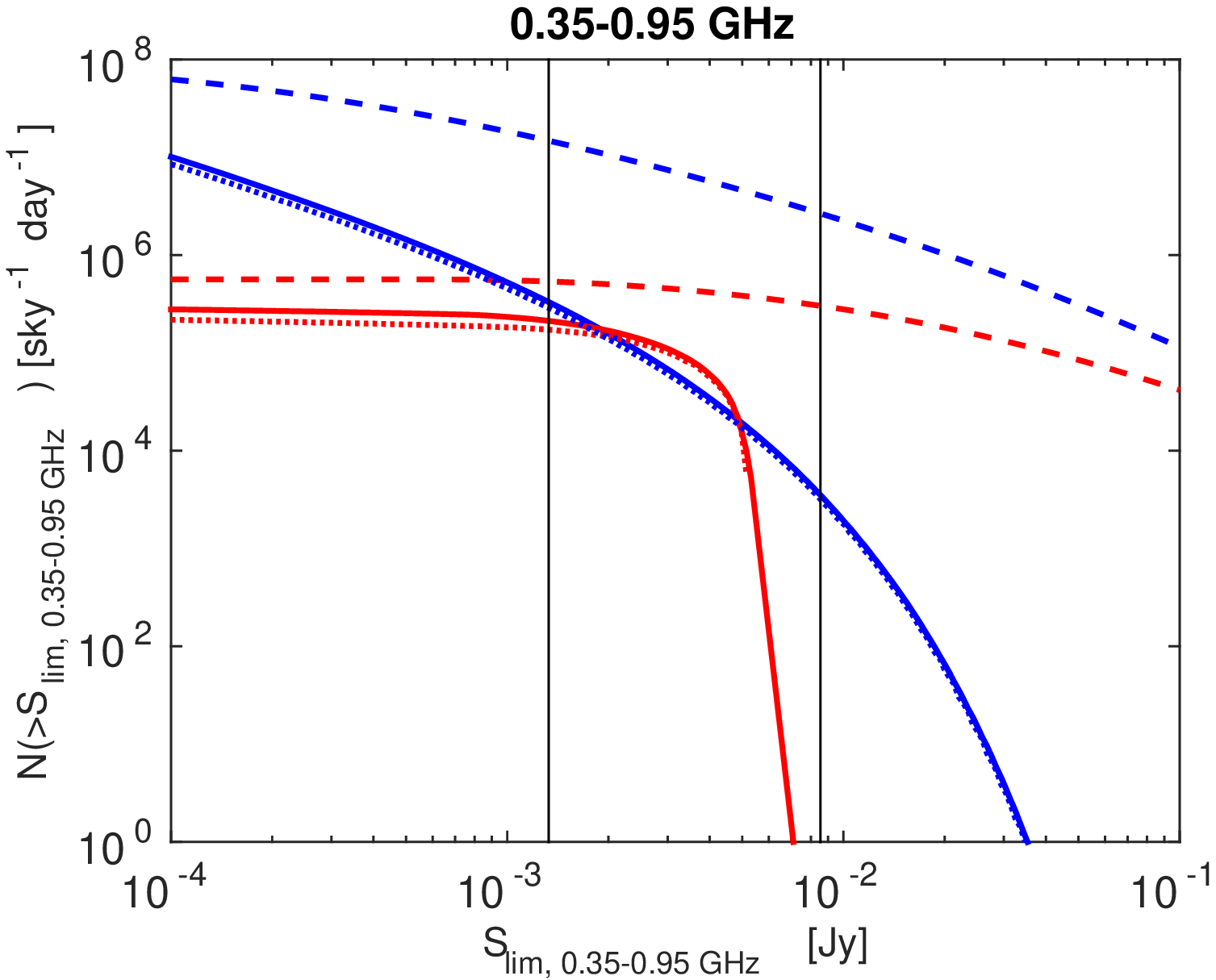}\includegraphics[width=2.2in]{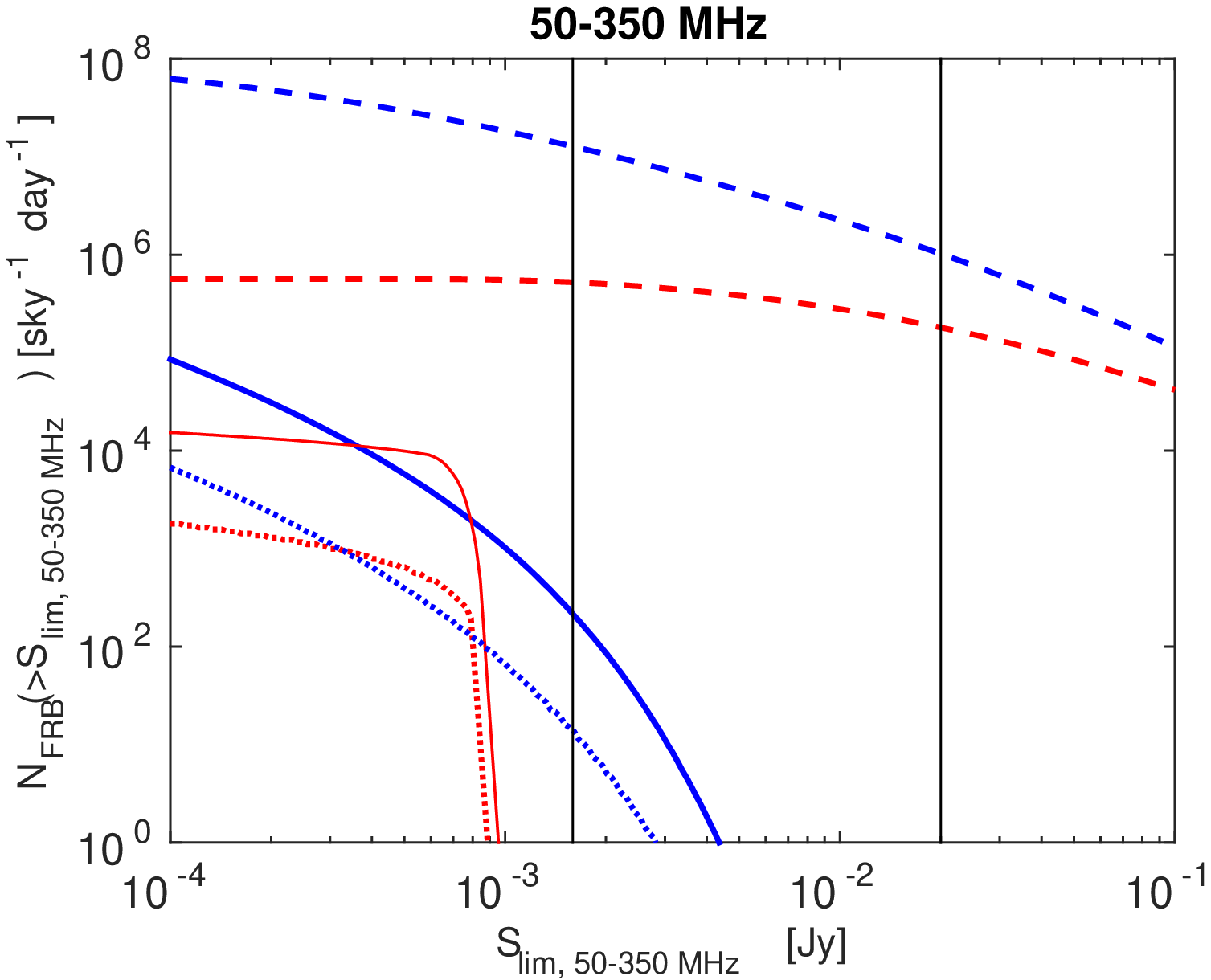}
\includegraphics[width=2.2in]{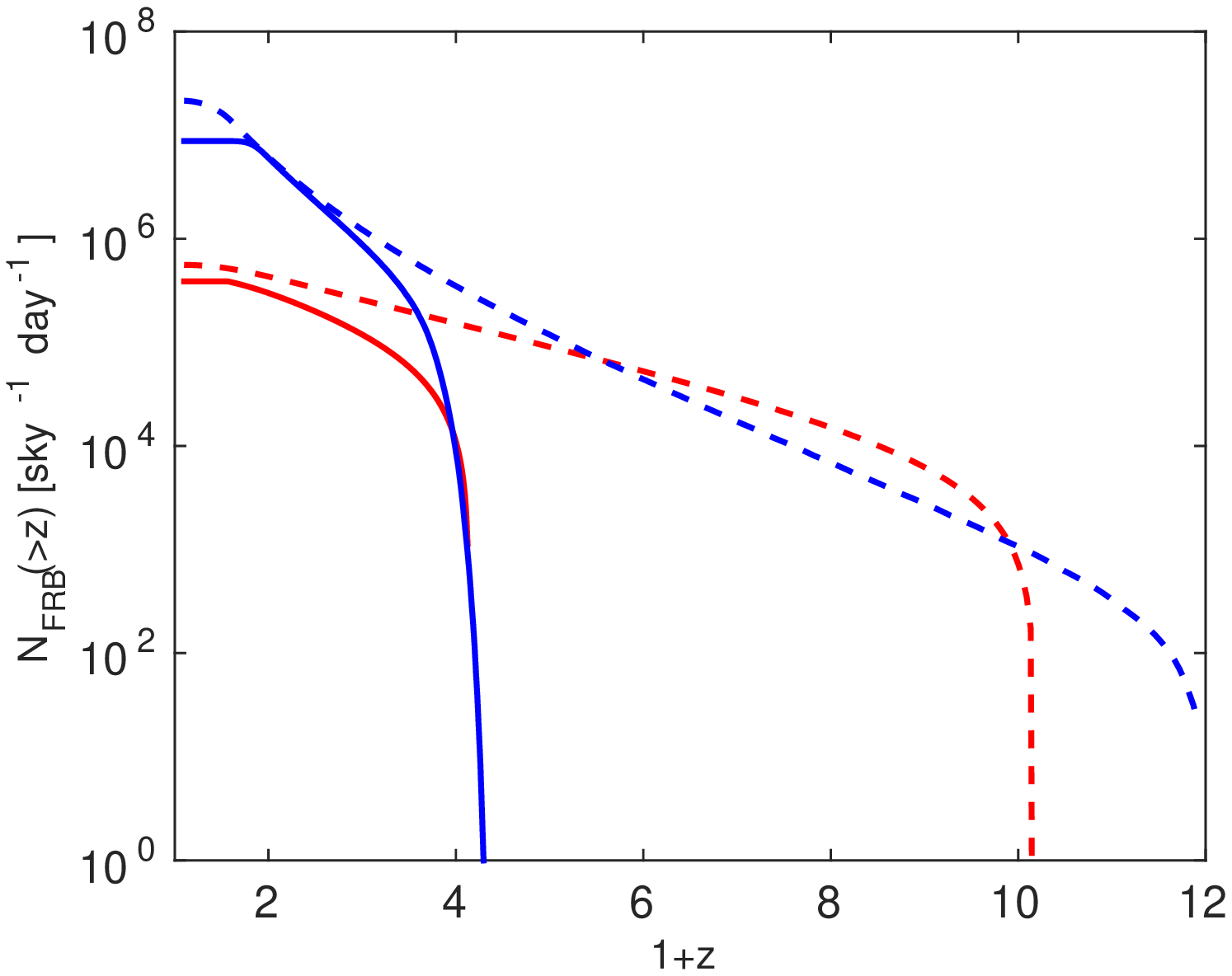}\includegraphics[width=2.2in]{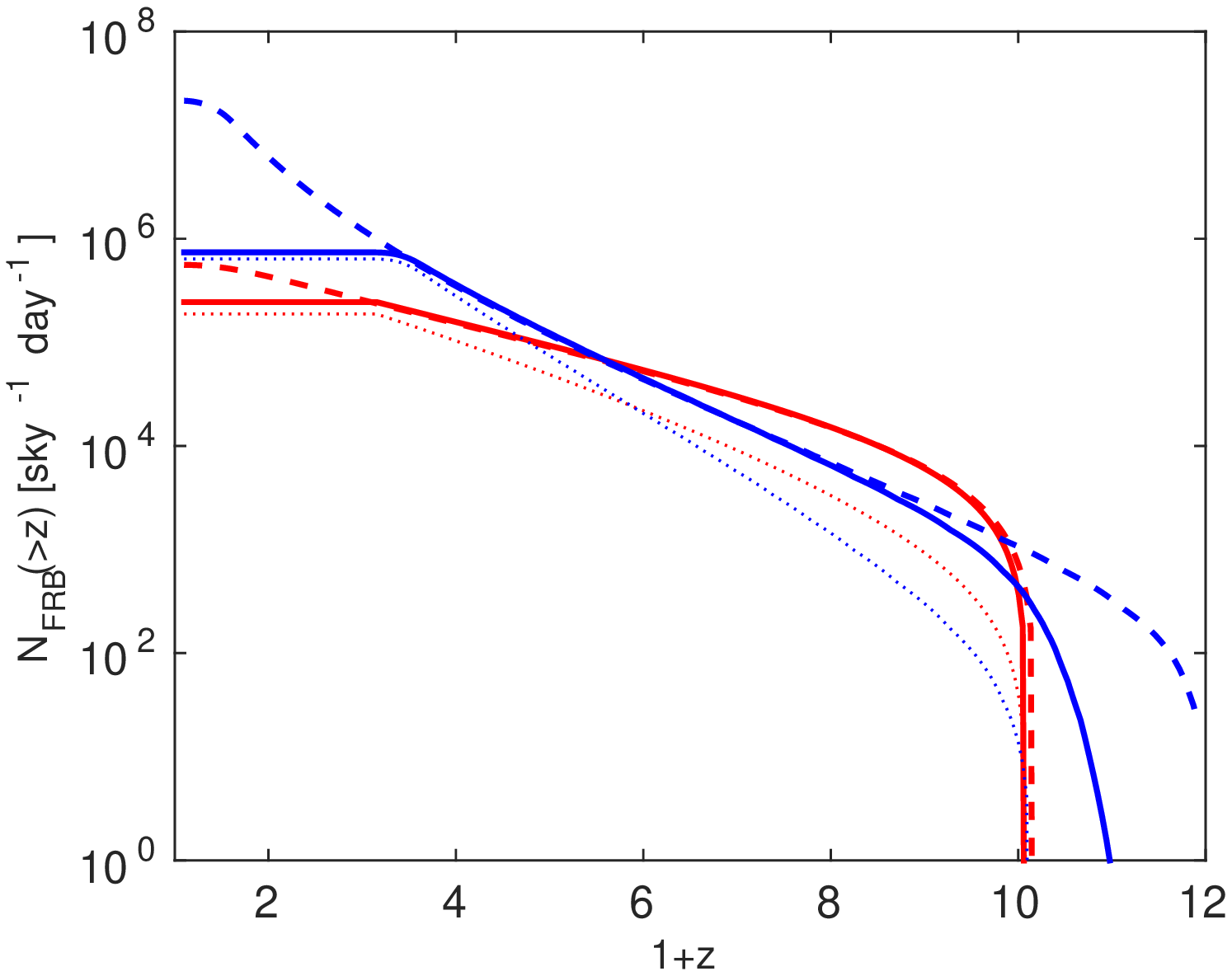}\includegraphics[width=2.2in]{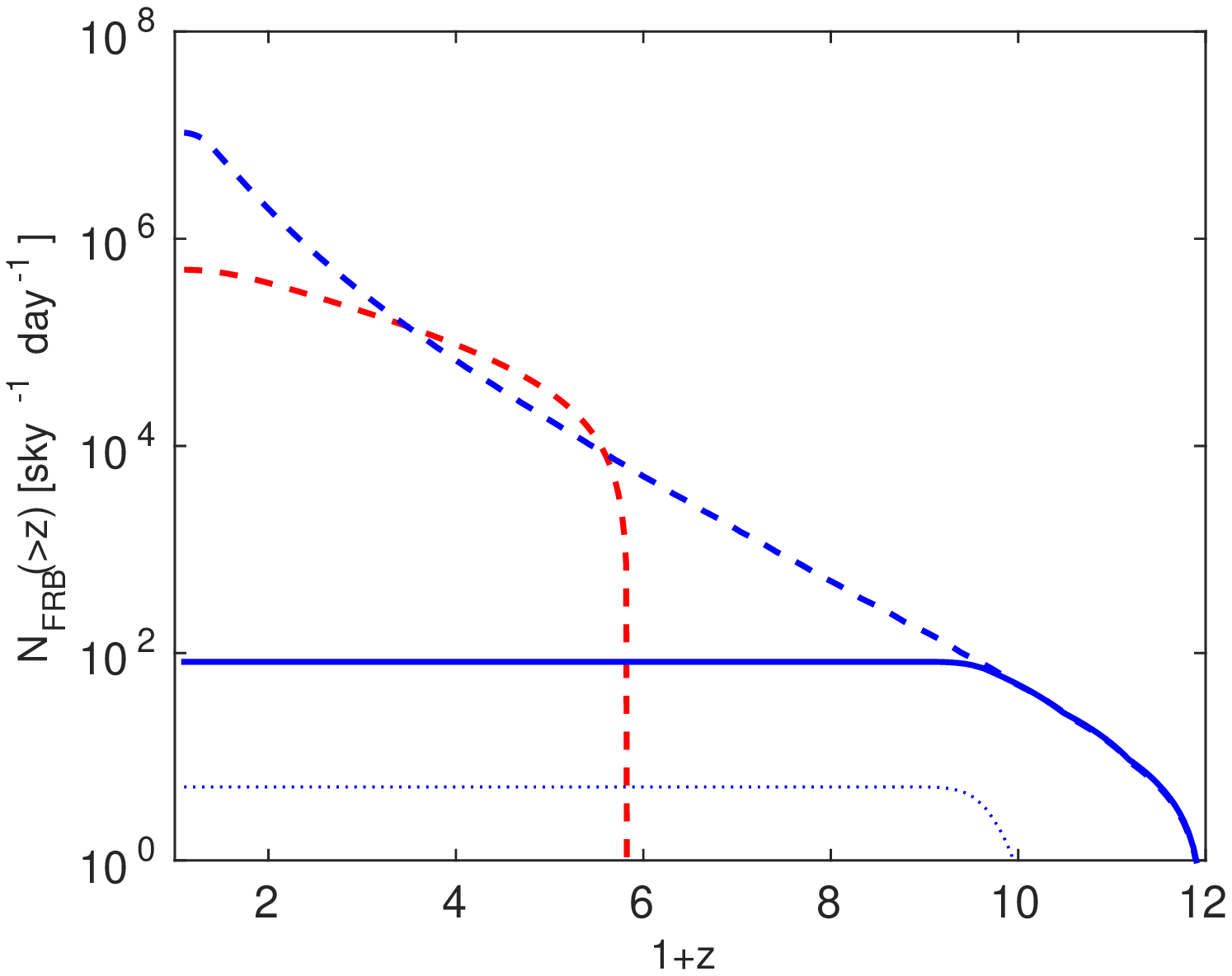}
\end{center}
\caption{ Detection rates of FRBs predicted for the  SKA-MID2 (left),  SKA-MID1 (middle) and the SKA-LOW (right) as a function of flux limit (top) and the minimal redshift of FRBs (bottom) detected by an observatory such as SKA (phase-2) with a flux limit of 0.8 mJy in the $0.95-1.76$ GHz band, 0.8 mJy in the $0.35-0.95$ GHz band and 2 mJy in the 50-350 MHz band.   We show the results for Models A (red solid),  B (red dashed), C (blue solid), and  D (blue dashed), assuming low-mass  host galaxies. For SKA-LOW and SKA-MID1 we also show the case of massive host galaxies (dotted lines).  On the top panels vertical lines indicate the $10-\sigma$ sensitivity limit for the phase 1 and 2 of the SKA (right and left  lines, respectively) to a pulse of intrinsic duration of 1 ms located at $z = 9$ (SKA-LOW), 2.7 (SKA-MID1) and 1 (SKA-MID2). }   
\label{Fig:LFska}
\end{figure*}

We apply the formalism discussed above to make predictions for the number of FRBs detectable with future observatories. We choose to focus our attention on the  Square Kilometer Array (SKA) as this facility will be very sensitive and  have two different low-frequency instruments thus being able to probe faint high-redshift FRBs.  The two arrays are: the low band array (SKA-LOW) at 50-350 MHz with sensitivity of 2 mJy and the high band instrument (SKA-MID) probing frequencies between $0.35-0.95$ GHz (SKA-MID1) and $0.95-1.76$ GHz (SKA-MID2) at a sensitivity of 0.8 mJy.  Detection rates of FRBs predicted for the SKA as a function of flux limit and the minimal redshift of FRBs are shown in  Figure \ref{Fig:LFska}. Another powerful machine to detect FRBs will be the Canadian Hydrogen Intensity Mapping Experiment (CHIME), which  is expected to have 125 mJy flux limit  in the $400-800$ MHz frequency range and a large collecting area \citep{Newburgh:2014, Rajwade:2017}.  Other relevant experiments (which, however, are not as sensitive as the SKA and have smaller field of view than CHIME) include, among others, the Low-Frequency Array  \citep[LOFAR][]{Coenen:2014} with 107 Jy sensitivity at 142 MHz, and the Green Bank Northern Celestial Cap (GBNCC) at 350 MHz with a sensitivity of 0.63 Jy to a 5 ms signal \citep{Chawla:2017}.

 We find that spectral shape of the bursts  is very important and strongly affects  the predictions for the high redshift and low frequency surveys. 
The top panels of Fig. \ref{Fig:LFska} shows the estimated FRB rates in the SKA bands. Models B and D (flat spectrum) yield band-independent predictions. A very sensitive survey conducted at any frequency would measure the same number of sources out to the edge of the Universe (at least $\sim 7$ FRBs sky$^{-1}$ sec$^{-1}$ in total); while with its sensitivity limit of 125 mJy CHIME will observe at least $3\times 10^4$ events sky$^{-1}$ day$^{-1}$ \citep[in agreement with numbers reported in the literature, e.g., ][]{Rajwade:2017}. 

However, the expected rate is much lower for models with the Gaussian spectral shape (Models A and C), and in this case each one of the instruments will probe an FRB population from a separate redshift range defined by the frequency band. In particular, CHIME will target the redshift range of $3.4-7.9$; however, such events would be below the flux sensitivity limit of the instrument. But owing to its higher sensitivity, SKA will be able to detect the faint high redshift events.  Based on the central frequency of the spectrum, SKA-MID2 will probe FRBs from $z = 1-2.7$  which includes the peak of star formation activity \citep[e.g.,][]{Madau:2014}, SKA-MID1 will probe events from $z = 2.7-9$ thus including a large part of the epoch of reionization  \citep[EoR, e.g., ][]{Loeb} which extends between  $z\sim 6$ and $z\sim 12$ \citep[e.g., ][]{Planck}, and SKA-LOW will have the correct band to search for all the bursts above $z = 9$ occurring at the beginning of the EoR and during the cosmic dawn, an epoch preceding the EoR when first stars formed. As seen from the bottom panels of Figure  \ref{Fig:LFska}, because of the width of the spectral profile the FRB rates do not drop abruptly at the cutoff redshifts specified above.

Sensitivity of a telescope to transients depends on the width of the pulse, $W$, which for FRBs depends on the redshift, intrinsic  properties of the source and broadening due to the finite sampling interval of the survey \citep{Rajwade:2017}. The limiting flux at the $10-\sigma$ detection threshold is given by 
\begin{equation}
S_{lim} = 10 \frac{SEFD}{\sqrt{2 W \Delta\nu}},
\end{equation}
where $SEFD$ is the system equivalent flux density,  $\Delta \nu$ is the bandwidth and factor 2 accounts for the two polarization channels. Assuming that the observed duration of the pulse is dominated by the redshifted intrinsic width,  we estimate $10-\sigma$ sensitivity at the lowest redshift detectable in each one of the SKA bands (vertical lines in Figure  \ref{Fig:LFska}). Adopting characteristic sensitivities for the final SKA configuration (phase-2, 0.8 mJy in the $0.95-1.76$ GHz band, 0.8 mJy in the $0.35-0.95$ GHz band and 2 mJy in the 50-350 MHz band) we see that  SKA-MID2 will detect FRBs out to $z\sim 3.2$ at a rate of $\sim 1000$ sky$^{-1}$ day$^{-1}$ or higher.  SKA-MID1 will detect FRBs out to $z\sim 9.1$ at a rate above 100 sky$^{-1}$ day$^{-1}$, with $\sim 2\times 10^4$ sky$^{-1}$ day$^{-1}$ FRBs expected from the EoR ($z>6$) for low-mass galactic hosts and $\sim 6000$ sky$^{-1}$ day$^{-1}$ if the hosts are massive. Finally, SKA-LOW will only be sensitive to FRBs out to $z\sim11$ (and only if the LF has a Schechter form) when the rate drops below 1 sky$^{-1}$ day$^{-1}$.

Low-frequency sensitive surveys such as SKA-LOW, can be used to identify the nature of the host galaxies if the spectral shape is not flat. In Models A and C we expect to find less FRBs at high redshifts in the model with massive hosts compared to the case of low-mass hosts.  This is simply because  massive galaxies are rare at higher redshifts and the population builds up at low redshifts. In the SKA-MID bands, where the signal is dominated by the events at low redshifts close to  the peak of star formation,  it is difficult to distinguish between the predictions for  low-mass hosts and massive hosts.

\section{Conclusions}
\label{Sc:Conc}

Adapting a phenomenological approach we built a simple model of a cosmological population of FRBs and made predictions for the expected number counts of FRBs in sensitive surveys of future telescopes.  However, because the number of observed FRBs is still very small, it is extremely difficult to make strong statistical inference.  Our models aim to bracket expected rates of faint FRBs.  

We find that when counting FRBs, and in particular when aiming to observe either intrinsically faint or high-redshift sources, a few factors play an important role: \\
{\bf The spectral shape of individual FRBs.} Events with a  Gaussian-like profile, such as  FRB 121102, and a limited bandwidth would  quickly redshift out of the observed band in addition to the scaling of their flux with luminosity distance.\\
{\bf Luminosity function.} Compared to a Schechter LF, the standard candles case has a much lower fraction of faint events.   \\
{\bf Host galaxy population.} A population of massive galaxies builds up at lower redshifts compared to low-mass galaxies, such as the host of FRB 121102. Even when having similar numbers of bright FRBs at low redshifts, many more events are expected at high redshifts when most host galaxies have low masses and low metallicities. Thus the   evolution of the number counts with redhsift can identify the characteristic mass of the hosts.

Naturally, the expected number counts depend on both the properties of the FRB population and the sensitivity of the radio observatories. We find that FRB detection rate with CHIME strongly depends on the spectral profile of the events. In the case of a Gaussian-like profile, the frequency range of the instrument includes FRBs from redshifts $3.4-7.9$, which are expected to be too faint to be detected by the instrument. For other spectral profiles (e.g., flat spectrum)  CHIME will be a powerful machine to detect FRBs. 

We find that it should be relatively easy to detect FRBs out to redshift $\sim 3$ with an observatory such as the SKA-MID2, which is predicted to observe at least one FRB per second over the entire sky.  Although detection of higher redshift events will be more difficult, a full sky survey with SKA-MID1 will collect more than $14$ FRBs  per minute originating in the low-mass galaxies at redshifts $z= 6-9$ during the EoR (and $\sim 4$ FRBs per minute if hosts are massive).  Since DM measures the column of ionized gas, FRBs occurring at so high redshifts could be used to constrain the reionization history and measure the total optical depth for the cosmic microwave background \citep{Fialkov:2016}. Because SKA-LOW is less sensitive than SKA-MID and there are fewer hosts populating the early Universe, detection of FRBs at even higher redshifts will be more difficult. 

Unresolved weak FRBs contribute to cosmological radio background. However, their contribution, which amounts to $\sim 0.1\mu$Jy, is much smaller than  other cosmological  components. For comparison, the  21 cm signal of neutral hydrogen \citep{Barkana:2016, Loeb} has intensity of few Jy.

\acknowledgments
We thank P. K. G. Williams, C. Law and D. Lorimer for stimulating discussions. This work was supported in part by the Black Hole Initiative, which is funded by a grant from the John Templeton Foundation.


\begin{thebibliography}{}
\bibitem[Adam et al. (2016)]{Planck} Planck Collaboration XLVII, 2016, A\&A, 596, 108
\bibitem[Bannister et al. (2017)]{Bannister:2017} 	Bannister et al., 2017, ApJ Letters. arXiv: 1705.07581
\bibitem[Barkana et al. (2016)]{Barkana:2016} Barkana, R., 2016, PhR, 645, 1
\bibitem[Barkana \& Loeb (2000)]{Barkana:2000} Barkana, R., 2Loeb, A., 2000, ApJ, 531, 613

\bibitem[Bassa et al. (2017)]{Bassa:2017} Bassa, C. G., Tendulkar, S. P., Adams, E. A. K., Maddox, N., Bogdanov, S., et al., 2017, arXiv: 1705.0 

\bibitem[Behroozi et al. (2013)]{Behroozi:2013} Behroozi, P. S., Wechsler, R. H., Conroy, C., 2013, ApJ, 770, 57 

\bibitem[Beloborodov (2017)]{Beloborodov:2017} Beloborodov, A. M. 2017, arXiv:1702.08644
\bibitem[Burke-Spolaor \&  Bannister (2014)]{Burke-Spolaor:2014} Burke-Spolaor S., Bannister K. W., 2014, ApJ, 792, 19
\bibitem[Caleb et al. (2016)]{Caleb:2016} 	
Caleb, M., Flynn, C., Bailes, M., Barr, E. D., Hunstead, R. W., Keane, E. F., Ravi, V., van Straten, W., 2016, MNRAS, 458, 708

\bibitem[Caleb et al. (2017)]{Caleb:2017} Caleb, M. et al., 2017,  arXiv:1703.10173
\bibitem[Champion et al. (2016)]{Champion:2016} Champion D. J. et al. 2016, MNRAS, 460, L30

\bibitem[Chatterjee et al. (2017)]{Chatterjee:2017} Chatterjee, S., Law, C. J., Wharton, R. S., et al. 2017, Nature, 541, 58
\bibitem[Chawla et al. (2017)]{Chawla:2017} Chawla, P., Kaspi, V. M., Josephy, A., Rajwade, K. M., Lorimer, D. R., et al., 2017, ApJ, 844, 140 
\bibitem[Coenen et al. (2014)]{Coenen:2014} Coenen, T., van Leeuwen, J., Hessels, J., Stappers, B. W., Kondratiev, V. I., et al., 2014 A\&A, 570, 60
\bibitem[Cordes \& Lazio (2002)]{Cordes:2002} 
Cordes, J. M., Lazio, T. J. W., 2002, arXiv:0207156

	
\bibitem[Cordes \& Wasserman (2016)]{Cordes:2016} 
Cordes, J. M., Wasserman, I., 2016,MNRAS, 457, 232


\bibitem[Cordes et al. (2017)]{Cordes:2017} Cordes, J. M., Wasserman, I., Hessels, J. W. T., Lazio, T. J. W., Chatterjee, S., Wharton, R. S.,  
2017, ApJ, 842, 35


\bibitem[Fialkov \& Loeb (2016)]{Fialkov:2016} Fialkov A., Loeb A., 2016, JCAP, 05, 004
\bibitem[Hunt et al. (2016)]{Hunt:2016} Hunt, L., Dayal, P., Magrini, L., Ferrara, A., 2016, MNRAS, 463, 2002
\bibitem[Keane et al. (2011)]{Keane:2011}  Keane E. F., Kramer M., Lyne A. G., Stappers B. W., McLaughlin M. A., 2011, MNRAS, 415, 3065
\bibitem[Keane \& Petroff (2015)]{Keane:2015} Keane E. F., Petroff E., 2015, MNRAS, 447, 2858.
\bibitem[Keane et al. (2016)]{Keane:2016} Keane E. F. et al., 2016, Nature, 530, 453-456.
\bibitem[Lawrence et al. (2016)]{Lawrence:2016} Lawrence, E., Vander Wiel, S., Law, C. J., Burke Spolaor,
S., Bower, G. C. 2016, arXiv:1611.00458
\bibitem[Law et al. (2017)]{Law:2017}  Law, C. J., Abruzzo, M. W., Bassa, C. G., Bower, G. C., Burke-Spolaor, S., et al., 2017, arXiv:1705.07553

\bibitem[Loeb \& Furlanetto (2013)]{Loeb} Loeb, A.,  Furlanetto, S. 2013, The First Galaxies in the
Universe, Princeton University Press (Princeton)


\bibitem[Li et al. (2016)]{Li:2016} Li, L., Huang, Y., Zhang, Z., Li, D.,  Li, B. 2016, arXiv:1602.06099

\bibitem[Lorimer et al. (2007)]{Lorimer:2007} Lorimer D. R., Bailes M., McLaughlin M. A., Narkevic, D. J., Crawford F., 2007, Science, 318, 777
\bibitem[Lorimer et al. (2013)]{Lorimer:2013} Lorimer D. R., Karastergiou, A., McLaughlin, M. A., Johnston, S., 2013 MNRAS, 436, 5L

\bibitem[Lunnan et al. (2014)]{Lunnan:2014} Lunnan, R., Chornock, R., Berger, E., et al. 2014, ApJ,
787, 138
\bibitem[Ma et al. (2016)]{Ma:2016} 
Ma, X., Hopkins, P. F., Faucher-Giguere, C.-A., Zolman, N.,  Muratov, A. L.,  et al.,  2016, MNRAS, 456, 2140
\bibitem[Madau \& Dickinson (2014)]{Madau:2014} Madau, P., Dickinson, M., 2014, ARA\&A, 52, 415
\bibitem[Masui et al. (2015)]{Masui:2015} Masui K. et al., 2015, Nature, 528, 523

\bibitem[Metzger et al. (2017)]{Metzger:2017}Metzger, B. D., Berger, E., \& Margalit, B. 2017, arXiv:1701.02370

\bibitem[Modjaz et al. (2008)]{Modjaz:2008} Modjaz, M., Kewley, L., Kirshner, R. P., et al. 2008, AJ,
135, 1136


\bibitem[Newburgh et al. (2014)]{Newburgh:2014} 	Newburgh, L. B., Addison, G. E., Amiri, M., Bandura, K., Bond, J. R., et al. 2014, SPIE, 9145, 4


\bibitem[Nicholl et al. (2015)]{Nicholl:2015} Nicholl, M., Smartt, S. J., Jerkstrand, A., et al. 2015, MNRAS, 452, 3869

\bibitem[Nicholl et al. (2017)]{Nicholl:2017}  Nicholl, M., Williams, P. K. G., Berger, E., et al. 2017,
arXiv:1704.00022 
\bibitem[Petroff et al. (2015)]{Petroff:2015} Petroff E. et al., 2015, MNRAS, 447, 246
\bibitem[Petroff et al. (2017)]{Petroff:2017} 	Petroff, E. et al., 2017, arXiv:1705.02911
\bibitem[Piro \&  Burke-Spolaor (2017)]{Piro:2017} Piro, A., Burke-Spolaor S., 2017, ApJ, 841L, 30

\bibitem[Rajwade \&  Lorimer (2017)]{Rajwade:2017}  Rajwade, K. M., Lorimer, D. R., 2017, MNRAS, 465, 2286
\bibitem[Ravi et al. (2015)]{Ravi:2015}  Ravi V., Shannon R. M., Jameson A., 2015, ApJ, 799, L5
\bibitem[Ravi et al. (2016)]{Ravi:2016}	Ravi, V. and Shannon, R. M. et al., 2016, Science
\bibitem[Sheth \& Tormen (1999)]{Sheth:1999}   	Sheth, R. K., Tormen, G., 1999, MNRAS, 308, 119
\bibitem[Spitler et al. (2014)]{Spitler:2014} Spitler L. G. et al., 2014, ApJ, 790, 101
\bibitem[Spitler et al. (2016)]{Spitler:2016} Spitler, L. G., Scholz, P., Hessels, J. W. T., et al. 2016, Nature, 531, 202
\bibitem[Tendulkar et al. (2017)]{Tendulkar:2017} Tendulkar, S. P., Bassa, C. G., Cordes, J. M., et al. 2017, ApJL, 834, L7
\bibitem[Thornton et al. (2013)]{Thornton:2013} Thornton D. et al., 2013, Science, 341, 53
\bibitem[Vedantham et al. (2016)]{Vedantham:2016} Vedantham, H. K., Ravi, V., Hallinan, G., Shannon,
R. M. 2016, ApJ, 830, 75

 
 \end{thebibliography}
\end{document}